\documentclass[twoside,twocolumn,english]{revtex4-2}
\usepackage[T1]{fontenc}
\usepackage[latin9]{inputenc}
\usepackage{amsmath}
\usepackage{amssymb}
\usepackage{graphicx}
\usepackage{wasysym}
\usepackage{esint}

\makeatletter

\providecommand{\tabularnewline}{\\}

\usepackage{soul}
\usepackage{color}
\usepackage{babel}
\usepackage{calc}

\makeatother

\usepackage{babel}
\begin{document}
\title{Ground States of the Mean-Field Spin Glass with 3-Spin Couplings}
\author{Stefan Boettcher and Ginger E. Lau}
\affiliation{$^{1}$Department of Physics, Emory University, Atlanta, GA 30322,
USA}
\begin{abstract}
We use heuristic optimization methods in extensive computations to
determine with low systematic error ground state configurations of
the mean-field $p$-spin glass model with $p=3$. Here, all possible
triplets in a system of $N$ Ising spins are connected with a bond.
This model has been of recent interest, since it exhibits the ``overlap
gap condition'', which should make it prohibitive to find ground
states asymptotically with local search methods when compared, for
instance, with the $p=2$ case better-known as the Sherrington-Kirkpatrick
model (SK). Indeed, it proves more costly to find good approximations
for $p=3$ than for SK, even for our heuristic. Compared to SK, the
ground-state behavior for $p=3$ is quite distinct also in other ways.
For SK, finite-size corrections for large system sizes $N\to\infty$
of both, the ensemble average over ground state energy densities and
the width of their distribution, vary anomalously with non-integer
exponents. In the $p=3$ case here, the energy density and its distribution
appear to scale with $\ln N/N$ and $1/N$ corrections, respectively.
The distribution itself is consistent with a Gumbel form. Even more
stark is the contrast for the bond-diluted case, where SK has shown
previously a notable variation of the anomalous corrections exponent
with the bond density, while for $p=3$ no such variation is found
here. Hence, for the 3-spin model, all measured corrections scale
the same as for the random energy model (REM), corresponding to $p=\infty$.
This would suggest that all $p$-spin models with $p\geq3$ exhibit
the same ground-state corrections as in REM. 
\end{abstract}
\maketitle

\section{Introduction\label{sec:Introdcution}}

Although conceived as a model to represent the properties of disordered
magnetic materials \cite{Edwards75}, Ising spin glasses -- and in
particular, their mean-field versions such as the Sherrington-Kirkpatrick
model (SK) \cite{Sherrington75} -- have found manifold applications
in various areas of science and engineering \cite{RSB40}. One especially
vibrant area of research has been the study of heuristics to solve
NP-hard combinatorial problems, for which such a spin glass with its
Boolean degrees of freedom provides about the most concise formulation
of a complex, hard-to-solve model \cite{Mezard06}. As such, it is
essentially equivalent to the quadratic unconstrained binary optimization
problem (QUBO) \cite{Kochenberger2014} more commonly used in computer
sciences and operations research (although some care must be taken
with the QUBO formulation \cite{Boettcher19}). For instance, QUBO
has become a benchmark problem for the application of quantum computing
to potentially solve NP-hard problems efficiently, at least, since
the emergence of the D-Wave quantum computer \cite{McGeoch13,Albash14,Rosenberg2016},
which launched a slew of ``quantum-inspired'' solvers typically
relying on vastly improved parallel processing hardware, such as Simulated
Bifurcation \cite{Goto19,Goto21}, Digital Annealer \cite{Aramon2019},
etc.

It is mostly for theoretical reasons that generalizations have been
introduced extending SK towards higher-order interactions beyond the
basic Ising spin glass, in which two-body interactions are represented
by bonds that constrain the behavior of two mutually connected spins.
For example, our study here was initially motivated by recent results
on the ``overlap gap'' property \cite{Gamarnik21}, which makes
finding global minima with local search prohibitive, that was shown
to apply to the 3-spin model but not SK \cite{Alaoui20}. The $p$-spin
model features $p$-body terms coupling together all possible combinations
of $p$ spins with a collective bond. For any positive integer $p$
and system size $N$, their Hamiltonian is given by
\begin{equation}
H\left(\vec{\sigma}\right)=-\sum_{1\leq i_{1}<i_{2}<\ldots<i_{p}\leq N}J_{i_{1}i_{2}\ldots i_{p}}\sigma_{i_{1}}\sigma_{i_{2}}\ldots\sigma_{i_{p}},\label{eq:H3spin}
\end{equation}
where $\sigma_{i}=\pm1$ are classical Ising spin variables and the
bonds $J$ are drawn randomly from some distribution ${\cal P}(J)$
of zero mean and unit variance. Most notable, it provides a means
to interpolate between SK (which corresponds to $p=2$) and probably
the most simple nontrivial spin glass called the random energy model
(REM), attained for $p\to\infty$ \cite{derrida:80,derrida:81}. The
$p$-spin model also resembles the iconic $K$-SAT problem in computer
science \cite{G+J}, correspondingly connecting $K$ Boolean variables
in each logical clause.

Our numerical results here support a sharp distinction between SK
($p=2$) and $p\geq3$ in their ground state behavior. Focusing on
ground state energies of $H$ in Eq. (\ref{eq:H3spin}) and their
distribution over the ensemble with bimodal bonds $J/J_{0}=\pm1$
in units of 
\begin{equation}
J_{0}=\sqrt{\frac{2}{N^{p-1}p!}},\label{eq:J0}
\end{equation}
we find results that resemble those of REM, which would imply that
there is also a commonality for all $p\geq3$ in the low-temperature
limit. This finding is buttressed by studies of a bond-diluted version
of the $p=3$ model, which exhibits virtually indistinguishable results
asymptotically, unlike for SK, where finite size corrections were
found to depend strongly on the dilution \cite{Boettcher20,Wang22}.
The dramatic change in ground state behavior between $p=2$ and $p=3$
is all the more remarkable as it resembles a similar distinction in
the thermodynamic properties between SK, which has a finite-temperature
phase transition of 2nd order, and the family of $p$-spin models
for all $p\geq3$ having a finite-temperature phase transition that
is of random first order \cite{Gardner85}. The similarity of our
results for $p=3$ and those of REM ($p=\infty$) suggest a relatively
small distinction between models for all $p\geq3$, all essentially
behaving REM-like at $T=0$.

\begin{table}
\begin{tabular}{|r|r|r@{\extracolsep{0pt}.}l|r@{\extracolsep{0pt}.}l|c|r|r@{\extracolsep{0pt}.}l|}
\hline 
 & \multicolumn{1}{r}{} & \multicolumn{2}{c}{Complete} & \multicolumn{2}{c|}{($q=1$)} &  & \multicolumn{1}{r}{Diluted} & \multicolumn{2}{c|}{($q=0.25$)}\tabularnewline
\hline 
$N$ & $n_{I}$ & \multicolumn{2}{c|}{$\left\langle e_{0}\right\rangle _{N}$} & \multicolumn{2}{c|}{$\sigma\left(e_{0}\right)$} &  & $n_{I}$ & \multicolumn{2}{c|}{$\left\langle e_{0}\right\rangle _{N}/\sqrt{q}$}\tabularnewline
\hline 
\hline 
16 & 100000 & -0&6751(1) & 0&03030(9) &  & 10000 & -0&6695(5)\tabularnewline
\hline 
18 & 100000 & -0&6870(1) & 0&02807(8) &  & 10000 & -0&6826(5)\tabularnewline
\hline 
20 & 100000 & -0&6977(1) & 0&02565(8) &  & 10000 & -0&6935(4)\tabularnewline
\hline 
22 & 100000 & -0&7059(1) & 0&02396(7) &  & 10000 & -0&7026(4)\tabularnewline
\hline 
24 & 100000 & -0&7135(1) & 0&02227(7) &  & 10000 & -0&7107(4)\tabularnewline
\hline 
26 & 100000 & -0&7197(1) & 0&02095(6) &  & 10000 & -0&7181(3)\tabularnewline
\hline 
28 & 100000 & -0&72531(9) & 0&01966(6) &  & 10000 & -0&7241(3)\tabularnewline
\hline 
30 & 100000 & -0&73001(9) & 0&01857(6) &  & 10000 & -0&7288(3)\tabularnewline
\hline 
32 & 100000 & -0&73436(8) & 0&01751(5) &  & 10000 & -0&7331(2)\tabularnewline
\hline 
34 & 100000 & -0&73816(8) & 0&01670(5) &  &  & \multicolumn{2}{c|}{}\tabularnewline
\hline 
36 & 100000 & -0&74167(7) & 0&01585(5) &  & 100000 & -0&74063(8)\tabularnewline
\hline 
38 & 100000 & -0&74461(7) & 0&01508(5) &  &  & \multicolumn{2}{c|}{}\tabularnewline
\hline 
40 & 100000 & -0&74754(7) & 0&01438(4) &  &  & \multicolumn{2}{c|}{}\tabularnewline
\hline 
42 & 100000 & -0&75008(7) & 0&01381(4) &  &  & \multicolumn{2}{c|}{}\tabularnewline
\hline 
44 & 100000 & -0&75250(6) & 0&01318(4) &  & 100000 & -0&75172(6)\tabularnewline
\hline 
46 & 100000 & -0&75462(6) & 0&01270(4) &  &  & \multicolumn{2}{c|}{}\tabularnewline
\hline 
48 & 100000 & -0&75673(5) & 0&01218(4) &  &  & \multicolumn{2}{c|}{}\tabularnewline
\hline 
50 & 100000 & -0&75854(5) & 0&01174(3) &  &  & \multicolumn{2}{c|}{}\tabularnewline
\hline 
52 & 100000 & -0&76047(5) & 0&01138(4) &  & 56000 & -0&75996(8)\tabularnewline
\hline 
54 & 100000 & -0&76197(5) & 0&01096(3) &  &  & \multicolumn{2}{c|}{}\tabularnewline
\hline 
56 & 100000 & -0&76362(5) & 0&01063(3) &  &  & \multicolumn{2}{c|}{}\tabularnewline
\hline 
58 & 80000 & -0&76484(6) & 0&01019(4) &  &  & \multicolumn{2}{c|}{}\tabularnewline
\hline 
60 & 66000 & -0&76635(6) & 0&00997(4) &  & 22000 & -0&7661(1)\tabularnewline
\hline 
64 & 30000 & -0&76883(8) & 0&00937(5) &  & 16000 & -0&7686(1)\tabularnewline
\hline 
70 & 10000 & -0&7721(1) & 0&00855(9) &  & 10000 & -0&7717(1)\tabularnewline
\hline 
80 & 10000 & -0&7765(1) & 0&00764(8) &  & 10000 & -0&7764(1)\tabularnewline
\hline 
98 & 10000 & -0&7823(1) & 0&00621(6) &  & 10000 & -0&7822(1)\tabularnewline
\hline 
128 & 8000 & -0&7887(1) & 0&00480(5) &  & 3000 & -0&7882(1)\tabularnewline
\hline 
160 & 5000 & -0&7928(1) & 0&00377(5) &  & 2000 & -0&7928(1)\tabularnewline
\hline 
194 & 2800 & -0&7959(1) & 0&00313(6) &  &  & \multicolumn{2}{c|}{}\tabularnewline
\hline 
256 & 500 & -0&7997(1) & 0&0023(1) &  & 200 & -0&7996(2)\tabularnewline
\hline 
\end{tabular}

\caption{\label{tab:AllData}List of all data obtained for the 3-spin model
for average ground state energies $\left\langle e_{0}\right\rangle _{N}$
and its deviation $\sigma\left(e_{0}\right)$. For each system size
$N$, these observables were averaged over $n_{I}$ instances, drawn
at random from the ensemble, using the approximate optima reached
with the EO heuristic. Columns on the left pertain to the complete
(undiluted) system, while those further on the right concern systems
that are diluted such that only 25\% of the bonds are non-zero (bond
density $q=0.25$). }
\end{table}

\begin{figure}
\vspace{-0cm}

\hfill{}\includegraphics[viewport=10bp 90bp 650bp 525bp,clip,width=1\columnwidth]{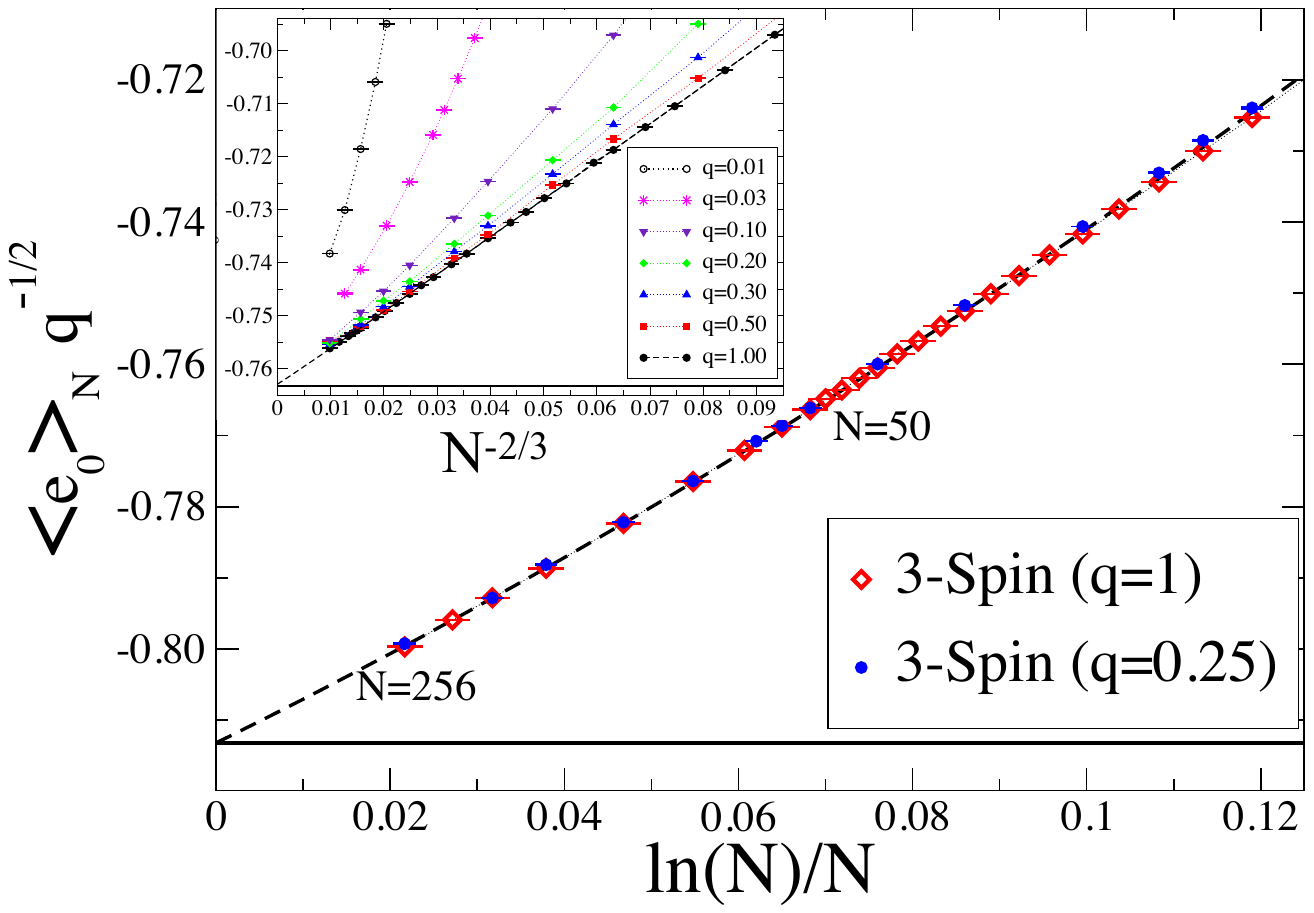}\hfill{}\vspace{0cm}

\caption{\label{fig:enerextra_3spin}Extrapolation plot for the average ground
state energy densities $\left\langle e_{0}\right\rangle _{N}$ as
a function of system size $N$. Shown is the data for the 3-spin model
for the completely connected case, i.e., having bond density $q=1$
(red $\lozenge$), and for a bond density of $q=0.25$ (blue $\CIRCLE$).
The most consistent form of finite size corrections, such that the
data extrapolates at the $y$-intercept to the known thermodynamic
limit at $N\to\infty$ of $\left\langle e_{0}\right\rangle _{N=\infty}=-0.8132$
(indicated by the horizontal line), proves to be $C_{N}=\ln N/N$,
with the dashed line providing a quadratic fit in this variable for
fixed $\left\langle e_{0}\right\rangle _{N=\infty}$. Inset: Corresponding
data for SK up to $N=1024$ for varying bond densities $q$, from
Ref. \cite{Boettcher20}, on a scale of $C_{N}=N^{-\frac{2}{3}}$.
Note the much more significant variation in corrections with $q$
in SK compared to the 3-spin model.}
\end{figure}

In Tab. \ref{tab:AllData} we have listed all our data for the ensemble
average of the ground state energies for both, the complete and the
diluted 3-spin system, as well as the deviation in the distribution
of ground state energies for the complete system. To analyze the data
for its asymptotic properties, we have first arranged the energies
$\left\langle e_{0}\right\rangle _{N}$ in an extrapolation plot according
to some presumed form of the finite-size corrections (FSC), 
\begin{equation}
\left\langle e_{0}\right\rangle _{N}\sim\left\langle e_{0}\right\rangle _{\infty}+AC_{N}+\ldots,\qquad\left(N\to\infty\right),\label{eq:FSC}
\end{equation}
where $C_{N}$ is the expected correction that should vanish for $N\to\infty$.
In SK, it is generally believed that $C_{N}$ has a power-law form,
$C_{N}\sim N^{-\omega}$, and several studies have conjectured that
$\omega=\frac{2}{3}$ for the corrections exponent \cite{EOSK,Aspelmeier07,Boettcher10b},
for which the SK ground state data, when plotted according to the
FSC in Eq. (\ref{eq:FSC}), provides an excellent extrapolation to
the exactly-known Parisi energy, $\left\langle e_{0}\right\rangle _{\infty}^{{\rm SK}}=-0.76321\ldots$
in the thermodynamic limit (see inset of Fig. \ref{fig:enerextra_3spin}).
For the 3-spin model here, such a power-law fit to the data in Tab.
\ref{tab:AllData} also allows for an exceptionally close fit for
all $N\geq20$, with a fitted exponent that is within 1\% of $\omega=\frac{4}{5}$.
However, the extrapolation predicts a fitted value of $-0.8151(1)\ldots$
for the ground state energy density at $N=\infty$, which is unacceptable,
since it is some $20\sigma$ from the exactly-known value of $\left\langle e_{0}\right\rangle _{\infty}^{p=3}=-0.8132\ldots$,
based on 1-step replica calculations \cite{Montanari03,Yeo20}. Fixing
$\left\langle e_{0}\right\rangle _{\infty}^{p=3}$ in the power-law
fit does not produce a stable result anymore, even when higher-order
corrections are included. While it is, of course, impossible to exclude
more exotic forms for $C_{N}$, or the effect of higher-order terms,
the proximity of a power-law exponent quite close to unity suggests
a correction of the form 
\begin{equation}
C_{N}\sim\frac{\ln N}{N},\label{eq:C_N}
\end{equation}
as is found exactly in REM. Indeed, when we again fix $\left\langle e_{0}\right\rangle _{\infty}^{p=3}$
and now fit the data in Tab. \ref{tab:AllData} according to the FSC
in Eq. (\ref{eq:FSC}) asymptotically for large enough $N$, we do
obtain a stable fit, as is shown in Fig. \ref{fig:enerextra_3spin}.
There, we have even allowed for a $C_{N}^{2}$ correction (dashed
line) that closely fits the data for all $N\geq32$.

We have probed the 3-spin model also for another of the subtle properties
that FSC in the ground state energies in SK exhibit. In Ref. \cite{Boettcher20},
it was found that in a bond-diluted version of SK, where $q$ is the
fraction of non-zero bonds, the FSC exhibit a power-law dependence
with an exponent $\omega=\omega(q)$ that varies with $q$, as shown
in the inset of Fig. \ref{fig:enerextra_3spin}. Already cursory exploration
of the 3-spin model at $q=0.25$ alone, with the data listed in Tab.
\ref{tab:AllData} and also plotted in Fig. \ref{fig:enerextra_3spin},
strongly suggests that there is no such variation with $q$ here.
Unlike for SK, this data (appropriately rescaled by $1/\sqrt{q}$
\cite{Carmona06}) neatly collapses onto that of the undiluted version
($q=1$) for all but the smallest sizes $N$, showing that its FSC
are likely identical at leading order, as in Eq. (\ref{eq:C_N}).
Since this form of $C_{N}$ does not have an anomalous exponent that
would allow a variation with $q$ to begin with, the observed lack
of variation further speaks in favor of such a form.

\begin{figure}
\vspace{0cm}

\hfill{}\includegraphics[viewport=0bp 20bp 740bp 550bp,clip,width=1\columnwidth]{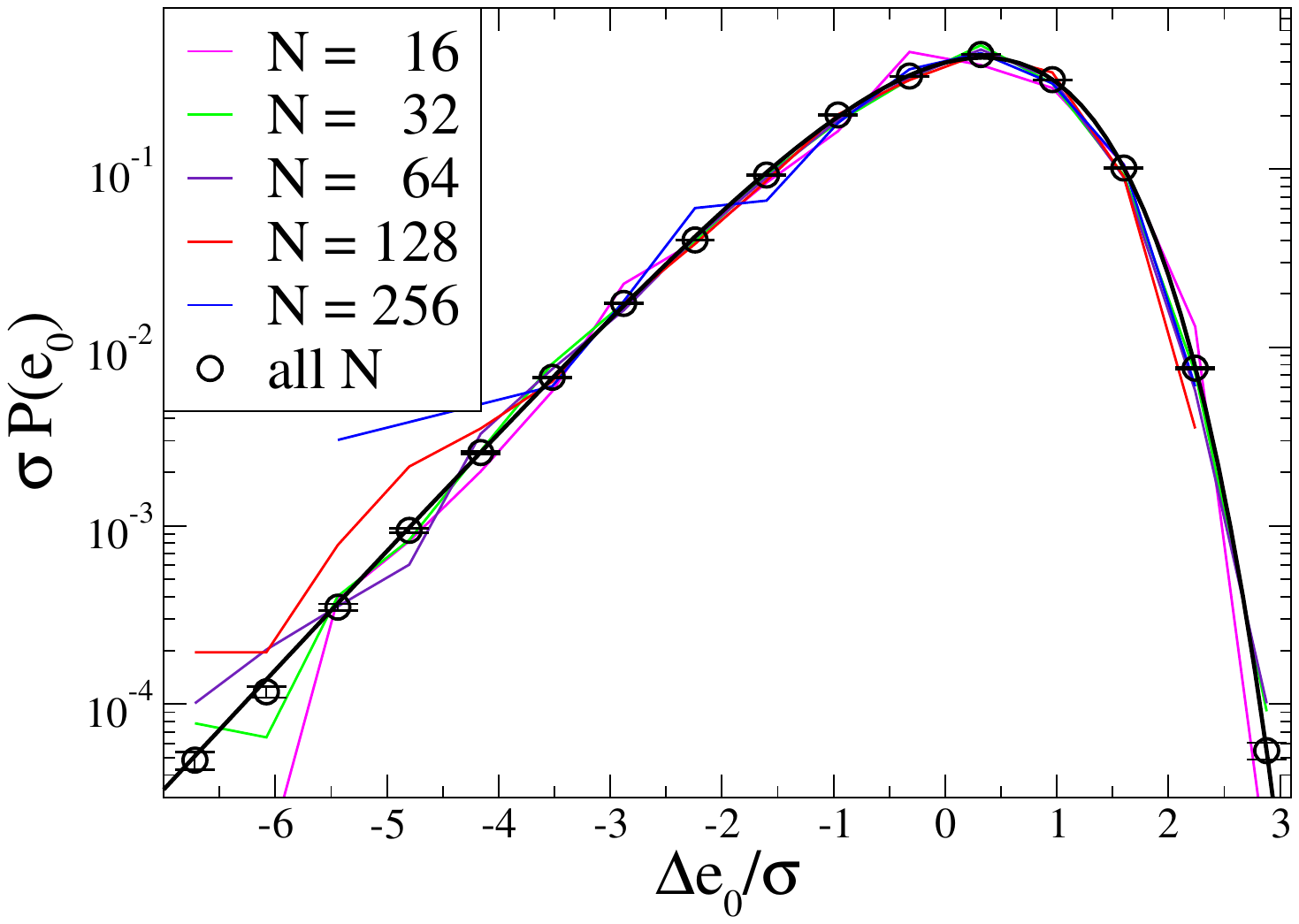}\hfill{}\vspace{0cm}

\caption{\label{fig:GSdistri3spin}Distribution of ground state energy density
fluctuations, $P\left(e_{0}\right)$, as a function of $\Delta e_{0}=e_{0}-\left\langle e_{0}\right\rangle $,
normalized by its deviation, $\sigma=\sqrt{\left\langle \Delta e_{0}^{2}\right\rangle }$.
Note that finite size effects are imperceptibly small, as indicated
by the colored thin lines for a range of system sizes $N$. Thus,
data for all sizes $N\protect\geq16$ have been combined ($\Circle$)
and fitted with a Gumbel distribution with a parameter of $m\approx0.87$
(thick solid line).}
\end{figure}

\begin{figure}
\vspace{0cm}

\hfill{}\includegraphics[viewport=0bp 20bp 740bp 550bp,clip,width=1\columnwidth]{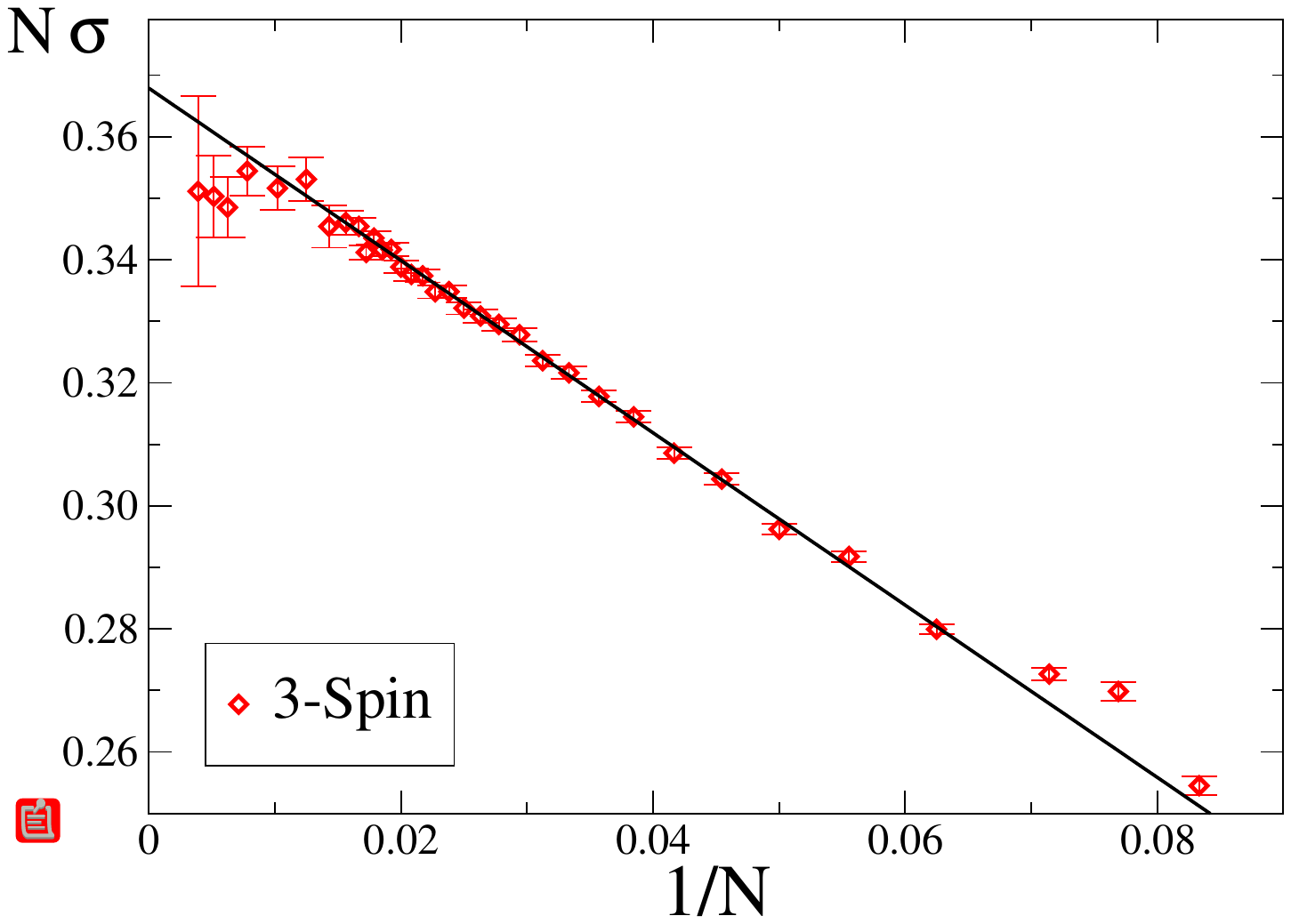}\hfill{}\vspace{0cm}

\caption{\label{fig:rho_extra_3spin}Extrapolation plot for the finite size
corrections of the deviation $\sigma$ of the distribution of ground
state energies in Fig. \ref{fig:GSdistri3spin}. The data suggests
that $\sigma$ is simply a series in $1/N$, although logarithmic
corrections can not be entirely excluded.}
\end{figure}

Considering the distribution of ground state energies over the ensemble,
$P\left(e_{0}\right)$, we find that it is rather close to a simple
Gumbel distribution expected for an extreme value statistic for the
lowest in a spectrum of \emph{iid} random energies, as can be calculated
exactly for REM. As shown in Fig. \ref{fig:GSdistri3spin}, there
its very little variation in the shape of that distribution, appropriately
normalized via its standard deviation, for the data shown. Thus, we
have combined all data for $16\leq N\leq256$ listed in Tab. \ref{tab:AllData}
to improve statistics especially in the exponentially small tails
of $P\left(e_{0}\right)$, resulting in a very stable set of data
points. Fitting those with a generalized Gumbel distribution in the
form of 
\begin{equation}
\ln P\left(e_{0}\right)=A+m\left[\left(x-u\right)/v-\exp\left\{ \left(x-u\right)/v\right\} \right],\label{eq:Gumbel}
\end{equation}
using the rescaled variable $x=\left(e_{0}-\left\langle e_{0}\right\rangle \right)/\sigma$
with deviations $\sigma\left(e_{0}\right)=\sqrt{\left\langle e_{0}^{2}\right\rangle -\left\langle e_{0}\right\rangle ^{2}}$,
yields $m=0.87(5)$, $A=0.50(4)$, $u=0.33(5)$, and $v=1.27(4)$.
Note the proximity of $m$ to unity, marking a pure Gumbel distribution,
as found in REM. A corresponding fit to the ground state data in SK
is far less skewed, with a parameter of $m\approx5$ \cite{EOSK},
far more distinct from REM. The deviation itself appears to vary with
a simple $1/N$ corrections, as demonstrated in Fig. \ref{fig:rho_extra_3spin},
potentially with logarithmic factors that are hard to discern at this
scale. This behavior of the 3-spin model, again, matches more closely
with REM (which has $\sigma\sim\ln N/N$) than SK, for which it was
derived \cite{Parisi08} that $\sigma\sim N^{-\frac{5}{6}}$, a rare
exact finite size result for SK. 

\begin{figure}
\vspace{0cm}

\hfill{}\includegraphics[viewport=0bp 20bp 740bp 550bp,clip,width=1\columnwidth]{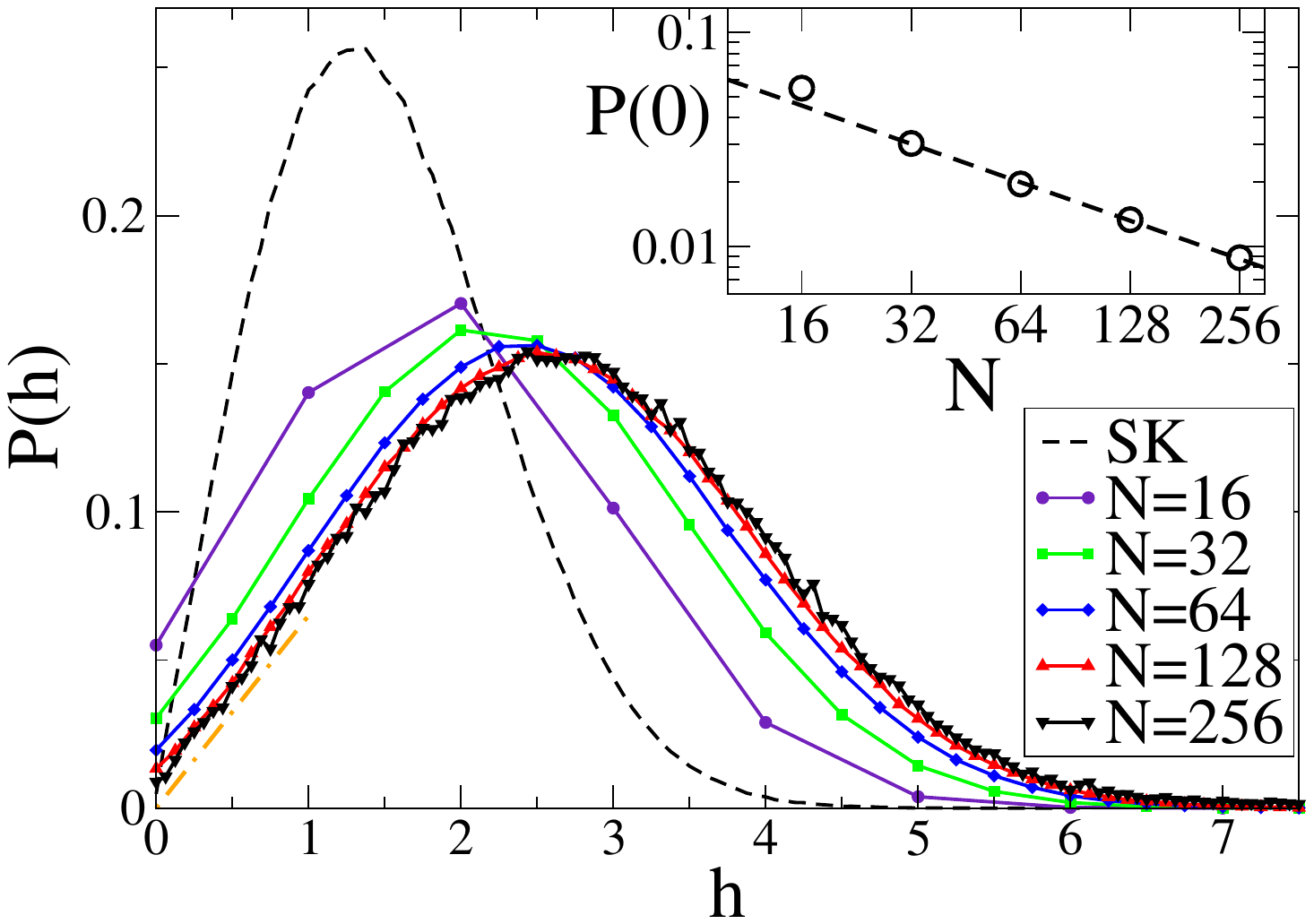}\hfill{}\vspace{0cm}

\caption{\label{fig:Ph3spin}Plot of the distribution $P(h)$ of local fields
$h$ impinging on spins by their neighbors in the ground state configurations
of the 3-spin model, plotted for different system sizes. (Since $P(h)$
is symmetric in $h$, only values for $h\protect\geq0$ are shown.)
Although significant finite-size effects are apparent, there is only
a small variation between the two largest sizes, suggesting that the
shape for $P(h)$ at $N=256$ is close to the thermodynamic limit.
For reference, the corresponding distribution for SK is provided (dashed
line). The orange dash-dotted line refers to Eq. (\ref{eq:P(0)}).
The values for $P(h=0)$ appear to evolve to zero, such that the distribution
has a pseudo gap in the thermodynamic limit. The inset suggests that
$P(0)\sim N^{-\alpha}$ for $N\to\infty$ with $\alpha\approx0.6$
(dashed line). }
\end{figure}

Finally, we have also measured the distribution $P(h)$ of local fields
$h_{i}$ that are impinging on each spin $\sigma_{i}$ due to the
coupling it has to all its neighbors in the ground state,
\begin{equation}
h_{i}=\sum_{j<h;i\not=j,k}J_{ijk}\sigma_{j}\sigma_{k}.\label{eq:localh}
\end{equation}
This distribution is shown for select values of $N$ in Fig. \ref{fig:Ph3spin}.
Its behavior is quite similar to that observed for SK at finite $N$
\cite{Boettcher07b}. The distribution is somewhat broader here but
similarly develops a pseudo-gap, i.e., $P(h)\to0$ linearly for $h\to0$,
in the thermodynamic limit. At finite size and near the origin, it
appears that we can make a linear fit,
\begin{equation}
P(h)\sim AN^{-\alpha}+Bh,\label{eq:P(0)}
\end{equation}
with parameters $\alpha\approx0.6$, $A\approx0.24$, and $B\approx0.06$
(see inset of Fig. \ref{fig:Ph3spin}).

\begin{figure}
\vspace{0cm}

\hfill{}\includegraphics[viewport=0bp 20bp 740bp 550bp,clip,width=1\columnwidth]{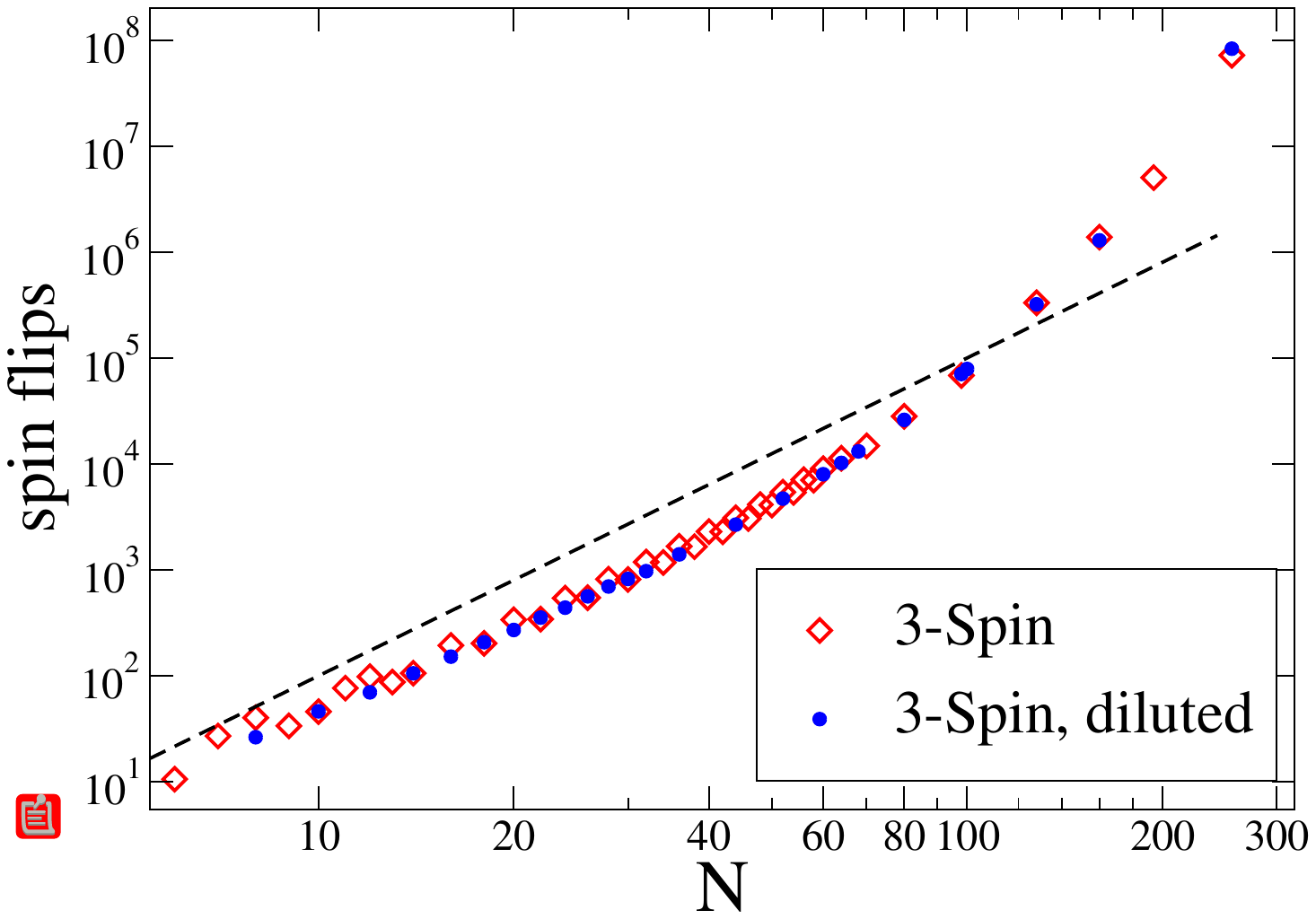}\hfill{}\vspace{0cm}

\caption{\label{fig:runtime3spin}Plot of the average number of spin flips
needed to reach the presumed ground state energy for each system size
$N$, both for the complete system (red $\lozenge$) and systems diluted
to 25\% of bonds (blue $\CIRCLE$). The maximal runtime is initially
set to be at least $0.1N^{3}$ spin flips (dashed line) but adaptively
increased such that always twice as many flips are used as were needed
to find the latest new optimum. With that criterion, runtimes escalate
rapidly for larger $N$ and become prohibitive above $N=256$.}
\end{figure}

To achieve our results for the $p$-spin model to this value in size,
$N=256$, we have implemented the extremal optimization heuristic
(EO) \cite{Boettcher01a,Boettcher00} similar to the version used
for SK in Ref. \cite{EOSK} in a vectorized form\footnote{The details of this implementation will be discussed elsewhere.}
on a GPU. To this end, we have also compressed both, the binary spin
variables as well as the discrete bimodal bond-distribution ($J=\pm1$
on one bit for the complete graph or $J=0,\pm1$ on two bits for the
dilute system), into a bit-wise form such that a single byte with
32-bits in the GPU can hold either 32 or 16 bonds. Although higher
system sizes than $N=256$ could have been reached with this implementation,
the necessary runtime to obtain stable results increased significantly
faster than for similar studies of SK \cite{EOSK}, as is shown in
Fig. \ref{fig:runtime3spin}. This might be caused by the fact that
local search in the $3$-spin model exhibits the ``overlap gap condition''
which entails a far more complex energy landscape to explore than
for SK \cite{Alaoui20}. Yet, using asymptotic extrapolations of our
finite-size data clearly makes a connection with the true thermodynamic
limit. Furthermore, the use of such an extrapolation plot (like Fig.
\ref{fig:enerextra_3spin}) of finite size data also proves to be
a powerful technique to assess the scalability of a heuristic, beyond
the use of static testbeds \cite{Boettcher19}, since low-$N$ ensemble
averages often yield sufficient bounds on the expected large-$N$
behavior to estimate systematic errors and detect their divergence
\cite{Boettcher22,Boettcher23}. 

In conclusion, we have provided the first systematic study of the
$T=0$ properties of the 3-spin model at finite size $N\leq256$ that
can serve as a benchmark for future comparisons. The results of our
extensive simulations for the ground state energies, an NP-hard problem
tackled with a sophisticated heuristic here, prove to be sufficiently
accurate at such large sizes that we can draw strong conclusions about
the nature of finite-size corrections by which the model extrapolates
to the thermodynamic limit. In the context of the generalized $p$-spin
model, we highlight the stark contrast in this behavior between the
case of $p=3$ here and the Sherrington-Kirkpatrick model ($p=2$)
on one side, and on the other side the close resemblance with the
REM ($p=\infty$), indicating how special SK is while all models with
$p\geq3$ appear to share the same finite size corrections to their
ground state behavior with REM. 

\paragraph*{Acknowledgements:}

We like to thank Ahmed El-Alaoui and Andrea Montanari for inspiration
and fruitful correspondences. SB thanks Federico Ricci-Tersenghi and
Lenka Zdeborova for helpful insights on the $p$-spin model. Some
of the computations were conducted on the Emory Hyper Community Cloud
Cluster (https://it.emory.edu/catalog/cloud-services/hyper-c3.html).

\bibliographystyle{apsrev4-2}
\bibliography{/Users/sboettc/Boettcher}

\end{document}